\documentclass[aps,prl,twocolumn,groupedaddress,showpacs]{revtex4}
\usepackage{graphicx}
\begin{document}
\title{Field- and pressure-induced phases in Sr$_{4}$Ru$_{3}$O$_{10}$: A spectroscopic investigation}
\author{Rajeev Gupta, M. Kim, H. Barath, S.L. Cooper}
\affiliation{Department of Physics and Frederick Seitz Materials Research Laboratory, University of Illinois,
Urbana, IL 61801}
\author{G. Cao}
\affiliation{Department of Physics, University of Kentucky, Lexington, KY 40506}
\date{\today}

\begin{abstract}
We have investigated the magnetic-field- and
pressure-induced structural and magnetic phases of the triple-layer
ruthenate Sr$_{4}$Ru$_{3}$O$_{10}$.  Magnetic-field-induced changes
in the phonon spectra  reveal dramatic spin-reorientation transitions and strong magneto-elastic
coupling in this material. Further, we are able to 
deduce key magnetoelastic coupling parameters, and evidence 
that the magnetic moments are localized on the Ru sites. Additionally, pressure-dependent Raman measurements at different 
temperatures reveal an anomalous negative Gruneisen-parameter associated with
the B$_{1g}$ mode ($\sim$ 380 cm$^{-1}$) at low temperatures (T $<$ 75K), which can be 
explained consistently with the field dependent Raman data.
\end{abstract}
\maketitle
\newcommand{\beq}{\begin{eqnarray}}
\newcommand{\eeq}{\end{eqnarray}}
\newcommand{\ve}{\varepsilon}
\newcommand{\bu}{{\bf u}}
\newcommand{\br}{{\bf r}}
\newcommand{\bq}{{\bf q}}
\newcommand{\bk}{{\bf k}}
\newcommand{\bcr}{{\bf R}}
\newcommand{\ex}{e^{i(kz-\omega t)}}
\newcommand{\ddt}{\frac{\partial^2}{\partial t^2}}

%
%
Spin-lattice coupling plays a critical role in the exotic 
properties and phases exhibited by a variety of oxide-based 
materials having both geometrical and chemical complexity; 
these include geometrically frustrated magnets such as 
ZnCr$_{2}$O$_{4}$ \cite{drew1}, low-dimensional spin systems 
such as  (TMTSF)$_{2}$PF$_{6}$ and CuGeO$_{3}$ \cite{lang}, 
hexagonal multiferroic manganites \cite{kimura}, and 
layered ruthenates \cite{karpus}.  The large spin-
phonon coupling in these materials leads not only to 
remarkable phenomena - such as magnetoferroelectricity, 
magnetic-field-induced metal-insulator transitions, 
and 'colossal' magnetoelastic effects - but also to highly-
tunable phase behavior in which structural properties can be 
sensitively manipulated by an applied magnetic field, or conversely in 
which magnetic properties can be controlled with applied 
pressure or strain.

The Sr-based layered ruthenates, such as double-layer 
Sr$_{3}$Ru$_{2}$O$_{7}$ and triple-layer Sr$_{4}$Ru$_{3}$O$_{10}$, are particularly 
interesting materials from the perspective of spin-lattice 
coupling.  For example, Sr$_{3}$Ru$_{2}$O$_{7}$ is an enhanced Pauli 
paramagnet \cite{ikeda,crawford} that exhibits induced 
ferromagnetism upon the application of hydrostatic pressure \cite{ikeda} 
or magnetic field \cite{perry,grigera}.  Density 
functional calculations suggest that the induced ferromagnetism in double-layer Sr$_{3}$Ru$_{2}$O$_{7}$ results from 
rotations of the RuO$_{6}$ octahedra, 
which lead to an orthorhombically distorted unit cell \cite{singh}.  Furthermore, triple-
layer Sr$_{4}$Ru$_{3}$O$_{10}$ is a structurally-distorted 
(antiferromagnetically) canted ferromagnet with a Curie 
temperature of T$_{C}$=105 K, in which the RuO$_{6}$ octahedra in the 
outer 2 RuO layers are rotated 5.25$^{\circ}$  about the 
c-axis, while the RuO$_{6}$ octahedra in the central RuO layer 
are rotated 10.6$^{\circ}$ about the c-axis in the 
opposite direction \cite{crawford}. Hence, one 
also expects  in this system a sensitive coupling between the 
structural and magnetic properties. For example, Raman measurements of Iliev et al. \cite{3} show that the frequency of the 380 cm$^{-1}$ B$_{1g}$ phonon in Sr$_{4}$Ru$_{3}$O$_{10}$ is highly sensitive to the onset of ferromagnetic order.

In this Letter, we investigate the intimate coupling 
between the spin and lattice degrees of freedom in Sr$_{4}$Ru$_{3}$O$_{10}$ 
using field- and pressure-dependent inelastic light 
scattering.  The strong spin-lattice coupling in the material  
allows us to use the frequency of the B$_{1g}$ phonon near 380 cm$^{-1}$ 
as a means of exploring the spin-spin correlation function 
as functions of field and pressure.  These 
results provide specific details regarding the strong spin-phonon coupling in Sr$_{4}$Ru$_{3}$O$_{10}$.  For example, we are  
able to deduce from these results the detailed H-T phase diagrams, the spin-phonon coupling parameter, and the specific magnetoelastic mechanism for the metamagnetic transition below 
T $<$ 50 K \cite{cao,mao} in Sr$_{4}$Ru$_{3}$O$_{10}$.

Our measurements were carried out on well-characterized single crystals (1 $\times$ 1 $\times$ 0.5mm$^{3}$) of Sr$_{4}$Ru$_{3}$O$_{10}$, which were grown by flux techniques \cite{cao}.  Transport and magnetization results on these samples are consistent with those in other published reports \cite{crawford,cao}.  The high purity and quality of the Sr$_{4}$Ru$_{3}$O$_{10}$ crystals studied here are clearly demonstrated by the observation of quantum oscillations,\cite{cao} as well as by the fact that these samples have a Dingle temperature that is less than 3 K, which is comparable to values found in high-quality organic crystals (0.5 to 3.5 K).  Field-dependent Raman measurements were performed in both Faraday (H $||$ c-axis $||$ $\vec{k}$) and Voigt (H $||$ ab-plane $\perp$ $\vec{k}$) configurations,\cite{karpus} and high-pressure Raman measurements were performed using a moissanite anvil cell \cite{snow}.

Magnetization \cite{crawford,cao} and x-ray diffraction measurements \cite{crawford} 
reveal that Sr$_{4}$Ru$_{3}$O$_{10}$ is a canted ferromagnet below T$_{C}$=105 K, 
with the Ru moments oriented primarily along the c-axis, as shown in Fig. 1(c).  
Notably, Fig. 1 shows that the temperature dependence of the B$_{1g}$ 
phonon frequency - which is associated with internal 
vibrations of the RuO$_{6}$ octahedra \cite{3} - exhibits a distinct change 
in slope, d$\omega$/dT, below T$_{C}$, as observed 
previously \cite{3}.  This anomalous frequency 
dependence is indicative of a strong magnetoelastic coupling 
between the B$_{1g}$ phonon mode and the c-axis ordered Ru-moments.

Upon the application of a magnetic field along the c-axis 
direction, the B$_{1g}$ phonon frequency exhibits a 
frequency increase with field, which has its largest value ($\sim$ 3 cm$^{-1}$) for T $<<$ T$_{C}$.  This field-induced phonon 
frequency shift below T$_{C}$ indicates that an applied field parallel to the c-
axis reduces the canting of the moments by causing the RuO$_{6}$ 
octahedra to increase their elongation along the c-axis, 
resulting in a contraction of the in-plane RuO bonds and a 
corresponding increase in the frequency of the in-plane B$_{1g}$ 
vibrational mode. This result identifies a specific structural mechanism associated with the weakly field-dependent c-axis magnetization observed previously in Sr$_{4}$Ru$_{3}$O$_{10}$ \cite{crawford,cao}. 

One can examine more quantitatively the magnetoelastic 
coupling between the RuO phonon and the Ru spins in the 
ferromagnetic (FM) phase by noting that the contribution of spin-
spin correlations to the phonon frequency can be approximated 
as \cite{5,6}, $\omega$=$\omega_{0}$ + $\lambda$ $\langle${\bf S}$_{i}$$\cdot${\bf S}$_{j}$$\rangle$, where $\omega_{0}$ is the bare phonon frequency in the absence of spin-phonon interactions,  
$\lambda$ is the spin-phonon coupling parameter, and ${\bf S}_{i}$ is the spin on the $i^{th}$ Ru site.  In Sr$_{4}$Ru$_{3}$O$_{10}$, 
the magnetic interactions are FM along the c-axis 
direction, and weak-antiferromagnetic (AFM) or paramagnetic (PM) in the ab-plane.  One 
can therefore approximate the spin-spin correlation function -
 within a molecular field approximation - by treating 
Sr$_{4}$Ru$_{3}$O$_{10}$ as a FM chain along the c-direction and 
taking an ensemble average over nearest-neighbor sites \cite{7}:
$\langle${\bf S}$_{i}$$\cdot${\bf S}$_{j}$$\rangle$ 
$\approx$ 2[M/2$\mu_B$]$^2$.  This relation assumes a 
saturation moment of 2 for the S=1 spin state of Ru.  
The resulting 
expression for the field-dependence of the magnetoelastic 
phonon frequency can be written:
\begin{equation}
\frac{\partial\omega}{\partial B} = 2\pi\lambda \left(\frac{M}{\mu_{B}^{2}}\right)\frac{\partial M}{\partial B}
\end{equation} 
Putting the measured values of $\frac{\partial\omega}{\partial B}$ =0.27 cm$^{-1}$/T (at 5K) and $\frac{\partial M}{\partial B}$ = 0.0083 $\mu_B$/T (at 1.7K \cite{cao}) in Eq. (1), we estimate the magnitude of the 
spin-phonon coupling constant for the B$_{1g}$ mode in Sr$_{4}$Ru$_{3}$O$_{10}$ to be $\lambda$ = 5.2 cm$^{-1}$ for T $<<$ T$_{C}$; this 
coupling reflects a strong sensitivity of the exchange 
interaction to atomic displacements below T$_{C}$, and is 
comparable to values of the 
coupling observed in other strong spin-lattice coupled 
systems such as  ZnCr$_{2}$O$_{4}$ \cite{drew1}. 

It is also of interest to examine the magnetoelastic effects of applying a magnetic field in the ab-plane direction. Previous magnetization measurements \cite{cao} with H $||$ ab-plane have identified a metamagnetic transition above H$_{c}$ $\sim$ 2 T for 
temperatures T $<$ 50 K, although the specific nature of this 
transition has not yet been identified.  Adding further 
interest to this transition, recent transport measurements of Sr$_{4}$Ru$_{3}$O$_{10}$
have identified abrupt resistive jumps in this field-induced 
transition, indicative of some form of switching behavior \cite{mao}.  In order to 
better identify the specific nature of this 
interesting field-induced transition, as well as the 
significance of the temperature scale T$^{*}$ below which this 
transition is observed, we show in Fig. 2 the field-
dependent Raman spectra of Sr$_{4}$Ru$_{3}$O$_{10}$ at various fixed 
temperatures for H $\vert \vert $ ab-plane.  Interestingly, Fig. 2 shows that there is little 
change in the B$_{1g}$ phonon frequency with increasing field for 
temperatures 50K $<$ T $<$ T$_{C}$.  However, for T $<$ 50 K, the B$_{1g}$ 
phonon frequency exhibits a significant decrease with 
increasing field up to a critical field H $<$ H$_{c}$ = 2 T, above 
which little or no additional change in the frequency is 
observed with field.  These results demonstrate a distinct 
structural contribution to the metamagnetic transition near 
H$_{c}$ $\sim$2 T, in which an applied field induces the Ru moments away from the c-axis, increasing the in-plane RuO bonds in the RuO$_{6}$ octahedra, and decreasing the B$_{1g}$ phonon frequency.  This implies that the metamagnetic transition for T $<$ T$^{*}$ $\sim$ 50 K and H$_c$ $\sim$ 2T is associated with a transition from an AFM or PM canted arrangement of the Ru moments to a FM canted arrangement (see Fig.3(a)). This result further suggests that the ``switching" behavior observed in transport measurements \cite{mao} through this field-induced transition is 
actually associated with inhomogeneous flipping of large regions of the sample from an AFM/PM canting arrangement to a FM canting arrangement.  The change from 
AFM/PM to FM canting most likely influences the in-plane conductivity by reducing in-plane 
spin-scattering. Note 
also that the strong magnetoelastic coupling evident in this 
material, which enables us to manipulate the structural 
properties with an applied magnetic field, suggests that the 
magnetic moments are localized on the Ru sites.

This interpretation of the metamagnetic transition in Sr$_{4}$Ru$_{3}$O$_{10}$ raises 
questions about the nature of the 
temperature scale T* $\sim$ 50 K below which the metamagnetic 
transition is observed.  We suggest that in the temperature 
range T* $<$ T $<$ T$_{C}$, the Ru moments are canted and 
FM aligned along the c-direction, but that 
there is no ordering of the in-plane component of the 
moments in the ab-plane.  For temperatures below T* = 50K, 
however, the moments ``lock" into an AFM 
canted or a PM configuration for H $<$ H$_{c}$ = 2T, and into a 
FM canted configuration for H $>$ H$_{c}$. 
We summarize the rich field-temperature phase diagram of 
Sr$_{4}$Ru$_{3}$O$_{10}$ for H $||$ ab-plane in Fig. 3(a).  Of particular 
interest in this phase diagram are the following features: at low temperatures
(T $<$ 50K), the system underdoes a metamagnetic transition from an AFM or PM canted 
configuration of the Ru moments to a FM 
canted configuration above H$_{c}$ $\sim$ 2T. We also show the variation of H$_{c}$ as a function of the in-plane applied field at different temperatures, which was obtained from isothermal magnetization measurements \cite{cao}. In the temperature range 50K $<$ T $<$ T$_{C}$, the Ru moments are FM aligned along the c-axis direction, but there is no net in-plane ordering for in-plane fields within our field range. For T $>$ T$_{C}$, the Ru moments are randomly oriented in all directions. 
For comparison, the simpler field-temperature phase diagram 
of Sr$_{4}$Ru$_{3}$O$_{10}$ for H $||$ c-axis is shown in Fig. 3(b).  The key 
feature of this phase diagram is a FM transition temperature that increases slightly from 105 K at 0T to $\sim$ 115 K at 
9T.  

Given its strong magnetoelastic coupling, it is of interest to investigate the extent to which one can influence the magnetic properties of Sr$_{4}$Ru$_{3}$O$_{10}$ by manipulating the structure at high pressures.  Fig. 4 summarizes the B$_{1g}$ phonon frequency as a function of temperature at various (quasi-hydrostatic) pressures, while 
the inset of Fig. 4 summarizes the derivative of the B$_{1g}$ 
mode frequency with respect to pressure, $d\omega/dP$, at various 
temperatures.  Note that $d\omega/dP$ is related to the mode 
Gruneisen parameter, defined as $\gamma_{i}$ = $\frac{1}{\omega_{i}\chi_{T}}$ $\frac{d\omega_{i}}{dP}$, where $\omega_{i}$ is the frequency of the $i^{th}$ mode, $P$ is the pressure, and $\chi_{T}$ is the isothermal 
compressibility.  Below T $\sim$ 70 K, there is a large decrease in 
the B$_{1g}$ phonon frequency with increasing pressure: 
$d\omega/dP$ = -0.32 cm$^{-1}$/kbar at T=5 K.  This is associated with the 
pressure-induced buckling of the RuO$_{6}$ octahedra on adjacent 
RuO layers, which forces the Ru moments to cant toward the ab-
plane with increasing pressure, in opposition to the 
tendency for c-axis FM ordering at P=0.  Consequently, 
the pressure-induced frequency shift of the B$_{1g}$ mode 
provides a measure of the magnetoelastic energy stored by 
the Ru moments and the RuO$_{6}$ octahedra in the FM phase.  
Interestingly, Fig. 4(a) shows that for P $\sim$ 24 kbar, the B$_{1g}$ 
phonon frequency has a roughly linear temperature 
dependence, with no anomalous change in slope through T$_{C}$, 
suggesting that the pressure-induced buckling of the RuO$_{6}$ 
octahedra on adjacent RuO layers at this pressure completely 
suppresses c-axis ferromagnetism for isobaric temperature sweeps, presumably by suppressing the RuO$_{6}$ rotations that 
accompany FM ordering in Sr$_{4}$Ru$_{3}$O$_{10}$.  
This conclusion supports the observation that the 
phonon frequency reduction between P=0 and P=24 kbar is $\Delta$ $\sim$ 7 cm$^{-1}$
at T=5K, which is comparable to the magnetoelastic energy estimated from the 
magnetic field dependent measurements.  Based upon these 
results, we infer the (P,T) phase diagram for Sr$_{4}$Ru$_{3}$O$_{10}$ 
shown in Fig. 4(b).  Our results suggest that high pressure magnetization measurements will 
exhibit a suppression of the c-axis magnetic moment for T $<$ 
T$_{C}$, as pressure causes increased canting of the Ru moments 
away from the c-axis direction.  Further, Fig. 4(b) shows that T$_{C}$ 
decreases with increasing pressure, culminating in the absence of 
c-axis ferromagnetism (for isobaric temperature sweeps) above P$^{*}$ $\sim$ 24 kbar, and that 
the ``lock-in" temperature T$^{*}$ increases with pressure.
  
In summary, we have used 
field- and pressure-dependent Raman spectroscopy to explore 
the strong spin-lattice coupling in Sr$_{4}$Ru$_{3}$O$_{10}$, and to 
provide a comprehensive microscopic description of the 
structural and magnetic (H,T) and (P,T) phases for this 
material.  In addition to a large spin-phonon coupling 
parameter $\lambda$ $\sim$ 5.2 cm$^{-1}$, we find evidence for a field-
induced change from an AFM or PM canted arrangement of the Ru 
moments to a FM canted arrangement for H$_{c}$ $>$ 2T (H $||$ ab-plane) 
and a suppression of the c-axis aligned FM phase for pressures 
above roughly 24 kbar. This 
strong magnetoelastic coupling provides strong evidence that 
the magnetic moments in this material are localized on the 
Ru sites.

We acknowledge support   by the National Science
Foundation under Grants No. DMR02-44502 (S.L.C.) and No. DMR02-40813 (G.C.), and the Department of Energy under 
Grant No. DEFG02-91ER45439 (S.L.C).

\widetext
\newpage
\begin{figure}[tb]
\centerline{\includegraphics[width=15cm]{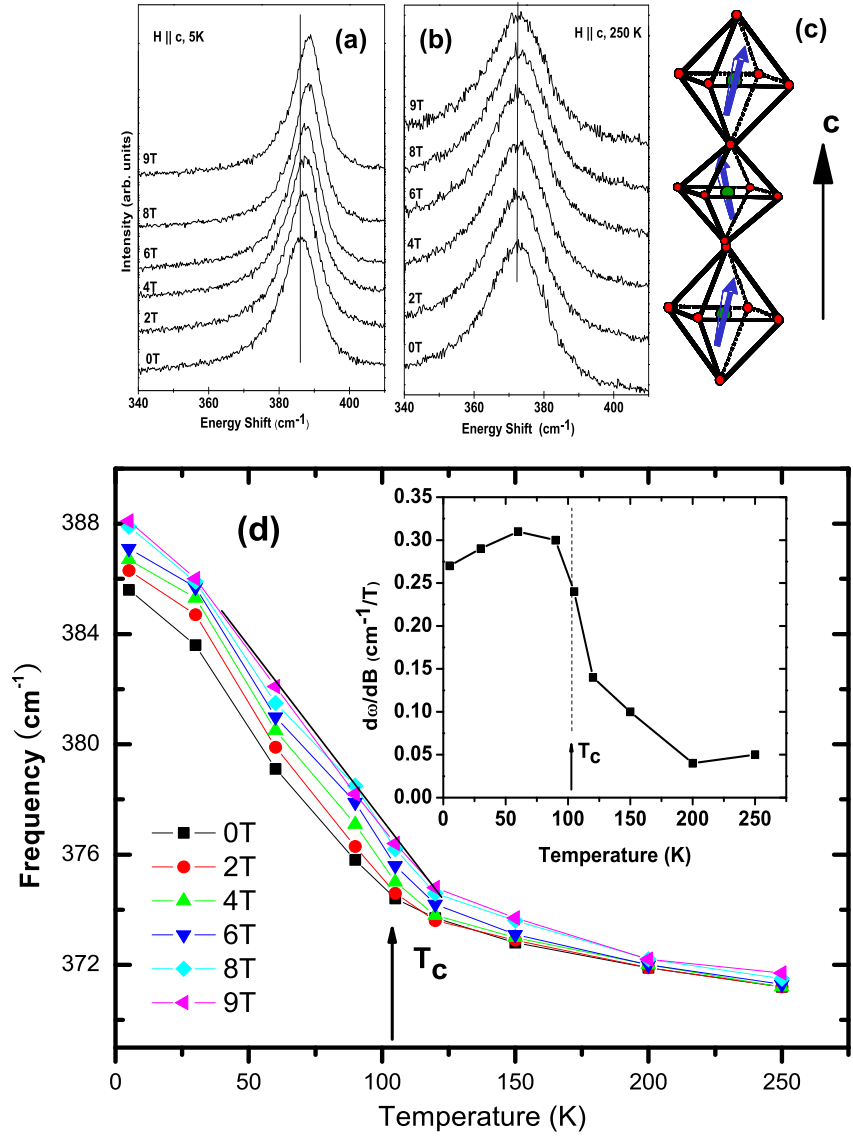}}
\caption{Representative Raman spectra of the B$_{1g}$ phonon mode in Sr$_{4}$Ru$_{3}$O$_{10}$ as a function of magnetic field with H $\vert \vert $ c-axis for (a) T=5 K and (b) T=250 K. (c) Picture 
of the orientation of the Ru moments in the FM phase in the 
three layers of octahedral RuO$_6$. (d) Temperature dependence of the phonon frequency ($\omega$) at different magnetic fields. The inset shows the slope $d\omega$/dB as a function of temperature.}
\end{figure}

\newpage
\begin{figure}[tb]
\centerline{\includegraphics[width=15cm]{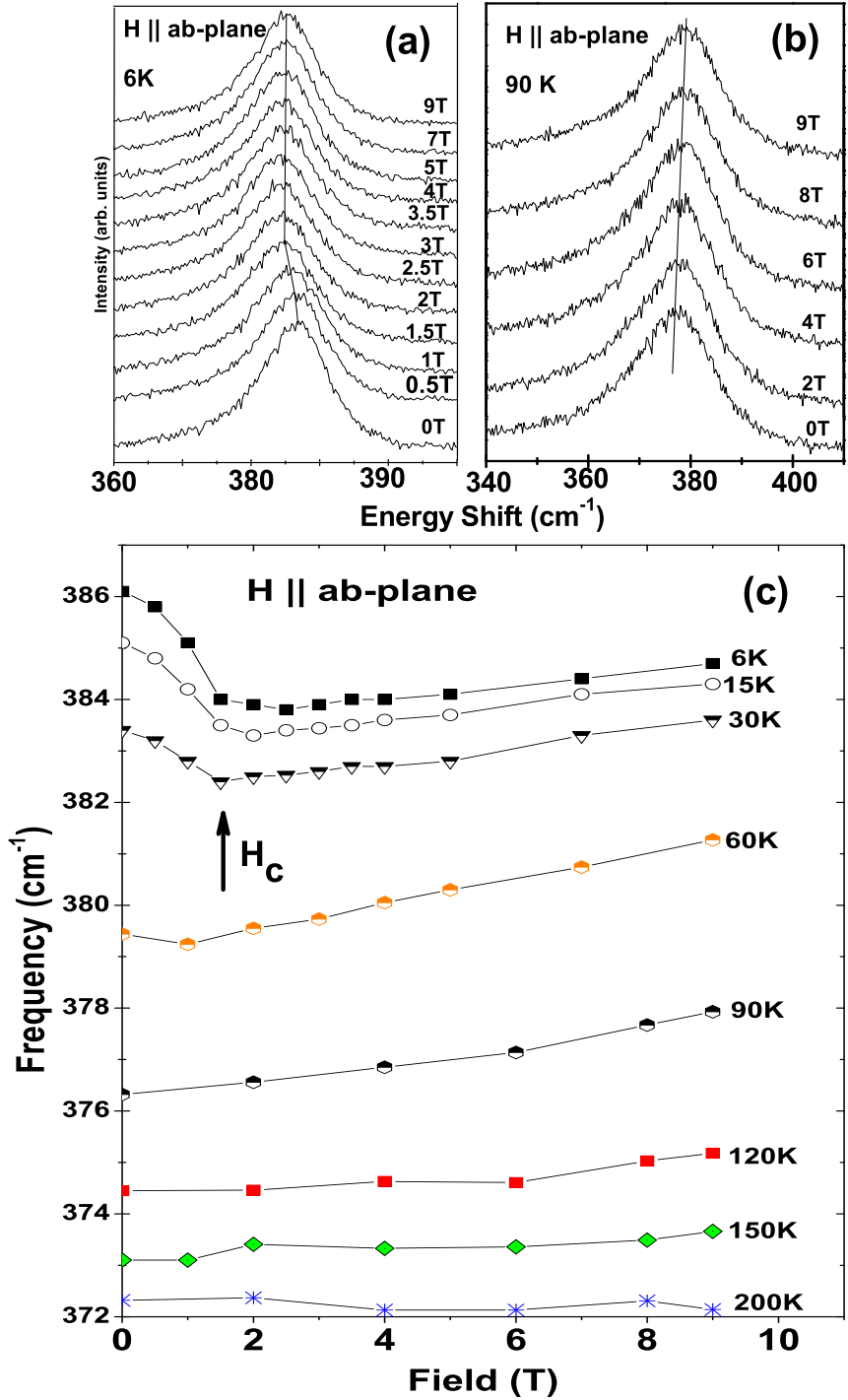}}
\caption{Raman spectra of the B$_{1g}$ phonon mode in Sr$_{4}$Ru$_{3}$O$_{10}$ as a function of magnetic field for H $\vert \vert $ ab-plane for (a) T=6 K and (b) T=90 K. (c) Field dependence of the B$_{1g}$ phonon frequency for different temperatures. Possible laser heating effects of about $\sim$ 5 K have not been included in the temperatures shown.}
\end{figure}

\newpage
\begin{figure}[tb]
\centerline{\includegraphics[width=15cm]{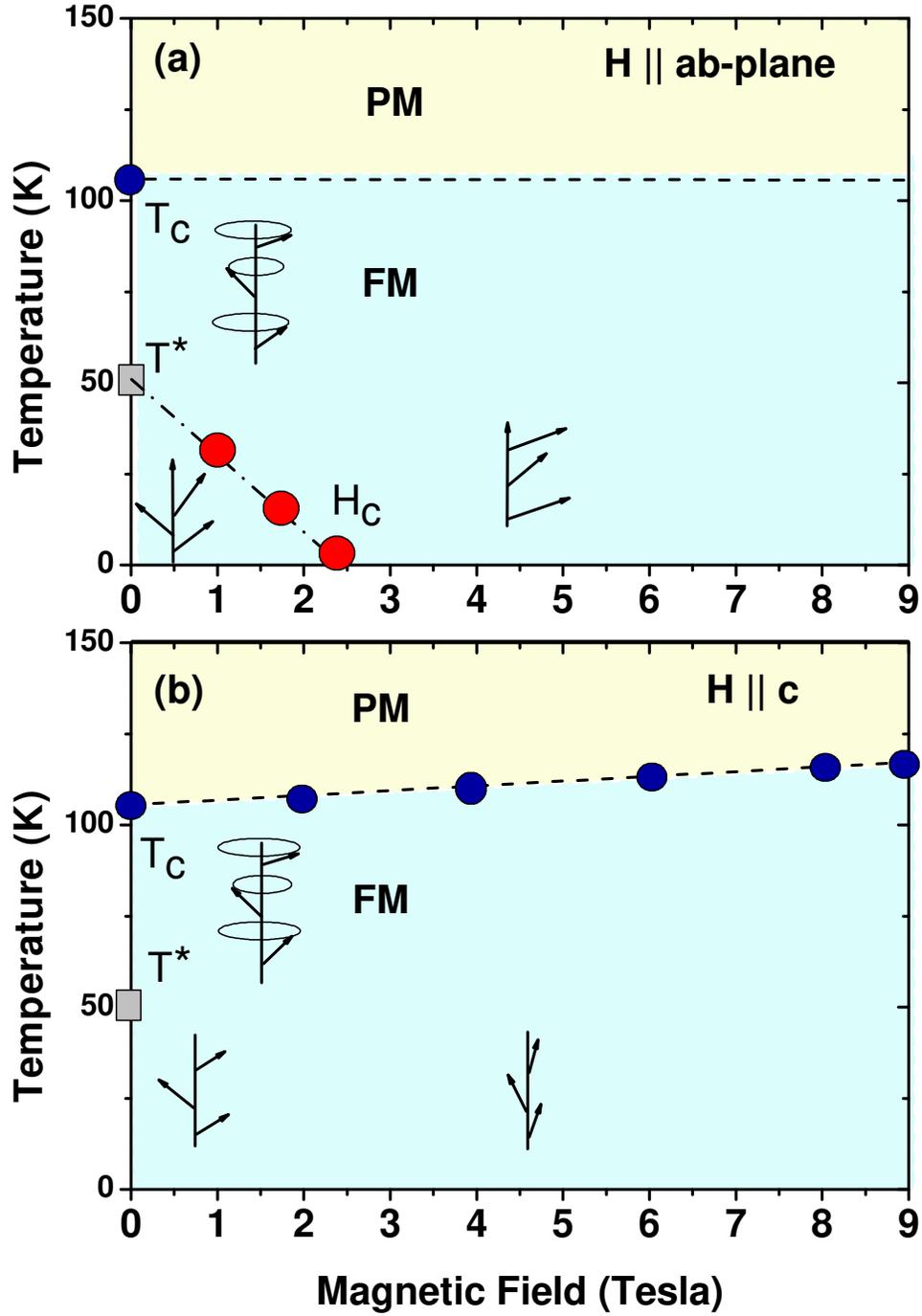}}
\caption{(a) Proposed (H,T) phase diagram 
for Sr$_{4}$Ru$_{3}$O$_{10}$ for H $||$ ab-plane, as deduced from the field-
dependent phonon data.  Red dots show the dependence of H$_{c}$ on temperature taken from magnetization measurements of ref.[11]. (b).  Proposed (H,T) phase 
diagram for Sr$_{4}$Ru$_{3}$O$_{10}$ for H $||$ c-axis. With increasing field, there is an increase of the Curie temperature, T$_{C}$, which is determined from the change in slope of B$_{1g}$ frequency, $d\omega/dT$.}
\end{figure}

\newpage
\begin{figure}[tb]
\centerline{\includegraphics[width=10cm]{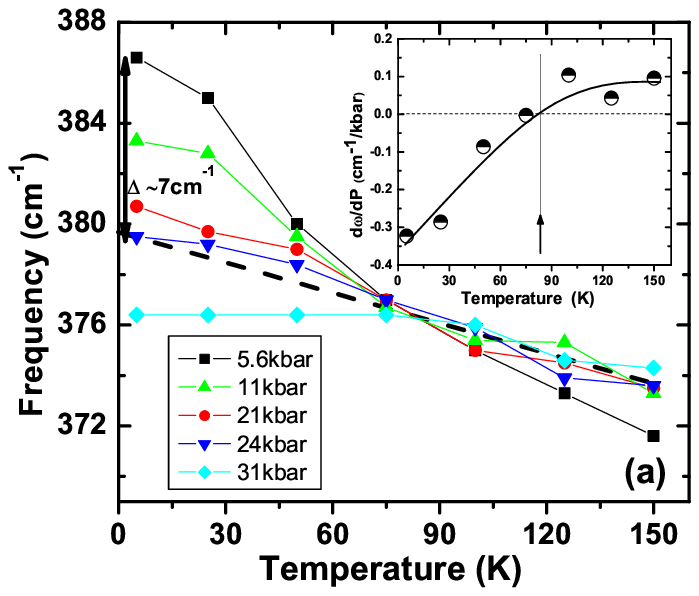}}
\centerline{\includegraphics[width=10cm]{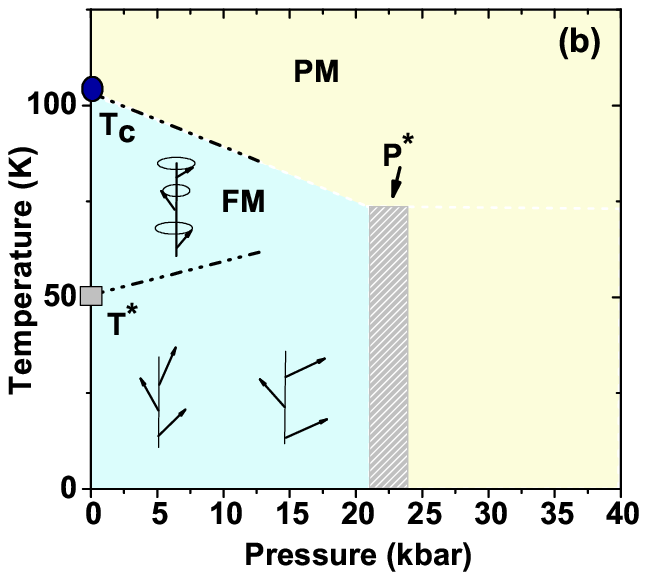}}
\caption{(a) Temperature dependence of the B$_{1g}$ phonon mode for different pressures. The inset shows the pressure derivative of the phonon frequency as a function of temperature, and the vertical line indicates the temperature ($\sim$ 75 K) below which $d\omega/dP$ has a negative value. (b) Proposed (P,T) phase diagram, for isobaric paths, for Sr$_{4}$Ru$_{3}$O$_{10}$, as deduced 
from the pressure-dependent phonon data.}
\end{figure}

\end{document}